\documentstyle{l-aa}

\begin{document}

   \thesaurus{ A\&A Supplement Series
                 (08.05.3, 08.08.1, 08.08.2, 08.16.3) } %

\title{Galactic Globular Cluster Stars: from  Theory to  Observation.}

\author{ S.Cassisi \inst{1}$^,$ \inst{2}$^,$ \inst{3}, V. Castellani \inst{4},
 S. Degl'Innocenti \inst{4}$^,$ \inst{5}, M. Salaris \inst{6}, A. Weiss \inst{6}}

   \offprints {S. Degl'Innocenti (Dipartimento di Fisica, Universit\`a 
di Pisa) scilla@astr18pi.difi.unipi.it}

\institute{
  Osservatorio Astronomico di Collurania, via Mentore Maggini, I-64100 Teramo, Italy
\and Dipartimento di Fisica, Universit\'a de L'Aquila, via Vetoio, I-67010 L'Aquila, Italy
\and Osservatorio Astronomico di Capodimonte, via Moiariello, I-80131 Napoli, Italy
\and Dipartimento di Fisica, Universit\'a di Pisa, piazza Torricelli 2,
   I-56126 Pisa, Italy 
\and Istituto Nazionale di Fisica Nucleare, Sezione di Ferrara, via Paradiso 12, I-44100 Ferrara, Italy
\and Max Plank Institut for Astrophysics, Karl Schwarzschild strasse 1, D-85470 Garching b. Munchen, Germany 
}

\date{Received ......... ; accepted ..........   }

\maketitle

\markboth {Cassisi et al.: Globular cluster stars}{Cassisi et al.: Globular cluster stars}

\begin{abstract}

We use evolutionary calculations presented in a recent paper 
(Cassisi et al. 1997a: hereinafter Paper I)
to predict B,V,I magnitudes for stars in galactic globulars. The effect of the
adopted mixing length on stellar magnitudes and colors is discussed, showing 
that the uncertainty on such a theoretical parameter prevents the 
use of MS stars as bona fide theoretical standard candles. 
However, comparison with Hipparcos data for field subdwarfs discloses  
a substantial agreement between theory and observation.
Present predictions concerning the magnitude of TO and of HB stars 
are compared with similar results appeared in the recent literature. 
We find that our predictions about the dependence
on metallicity of ZAHB magnitudes appear in good agreement with
observational constraints as recently discussed by Gratton et al. (1997c).
We present and discuss a theoretical calibration of the difference in 
magnitude between HB and TO as evaluated with or without 
element sedimentation.
The effect of a variation of the original helium content on the 
magnitude of MS, TO and HB stars is explored and discussed.  
Finally we use theoretical HB magnitudes to best fit the CM diagram 
of M68 and M5, taken as representative of metal poor and intermediate 
metallicity galactic globulars, deriving an age of 11$\pm$1.0 Gyr
 and 10$\pm$1.0 Gyr, respectively, for the adopted chemical compositions,
plus an additional uncertainty of $\pm$1.4 Gyr if the uncertainty on the
chemical composition is taken into account.
This result is discussed on the basis  of current evaluations
concerning cluster ages and distance moduli.

\end{abstract}

\keywords{Stars: evolution - HR diagram - horizontal-branch - Population II}


\section{Introduction}

Following the effort undertaken by Hubble Space Telescope  for  
improving the  determination  of the Hubble constant and by taking advantage 
of several improvements of the input physics a number of  authors
have recently revisited the prediction of stellar evolutionary
theory to re-evaluate the age of galactic globular clusters (see, e.g.,
VandenBerg et al. 1996; D'Antona et al. 1997; 
Salaris et al. 1997; Brocato et al.  1997; 
Chaboyer et al. 1998; Salaris \& Weiss 1998a,b).
In a previous paper (Cassisi et al. 1998: hereinafter Paper I) 
we discussed on  purely
theoretical grounds the effect of the improved physical inputs on 
stellar evolutionary models, evaluating the additional influence of element
diffusion on the evolutionary history of such  stars. In this
paper we  take advantage of such an homogeneous set of
stellar models to present theoretical predictions concerning 
observational magnitudes and colors, and to discuss the 
theoretical scenario leading to the evaluation of  cluster ages. 

Such age  evaluations, as primarily based on the calibration of the Turn-Off
luminosity in terms of the cluster age, requires the adoption
of suitable standard candles belonging to the cluster, 
as given either by MS or 
by HB stars. In the last case one is dealing 
with the so called $\Delta$V method, which is only formally 
independent of the cluster distance modulus. In all cases,
connecting theory to ages requires several steps worth being
discussed in some details. Thus, presenting the evolutionary results, 
we will also investigate the degree of reliability concerning
theoretical predictions about the magnitude of MS, TO and HB stars. 
The main  aim of this investigation is  to point out
theoretical  uncertainties but also to show how reliable the new models are,
testing the results with available observational constraints and, in
particular, with the absolute magnitudes of metal poor stars 
recently provided by  Hipparcos.

\section{From theory to observation}

In order to cover with sufficient details the range 
in metallicity Z=2$\cdot$10$^{-4}$ - 6$\cdot10^{-3}$ which 
appears  adequate for galactic globulars, computations presented in 
Paper I have been  supplemented
with an additional set of stellar models with Z=0.0006 .
As in Paper I, evolutionary models have been computed by adopting an 
amount of original He  Y=0.23 and with a mixing length 
parameter l = 1.6 Hp. Data for these
new models are reported in the Appendix of this paper.

On the basis of these new results, in this paper we will rely on a grid
of evolutionary models covering, at least, the range of cluster ages
8 -18 Gyr for the assumed metallicities Z= 0.0002, 0.0006, 0.001,
0.006. The computations have been performed both by including or
neglecting 
element sedimentation. In addition, for Z=0.001 we computed two set 
of evolutionary tracks with element sedimentation, under the same 
assumption about the mixing length but with Y= 0.21 or 0.25, respectively. 
These models will allow to discuss the influence of original He on models
where element diffusion is taken into account.
According to the usual procedure, the evolutionary models have been used 
to derive cluster isochrones in the logL, logTe plane.

\begin{figure}[htbp]
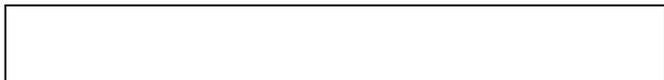
   
\picplace{1cm}
\caption{The run of the theoretical 12 Gyr isochrone in the 
V,B-V diagram for the labeled choices about the adopted model atmospheres.}
\label{atm}
\end{figure}

To allow a comparison with observation, this theoretical scenario 
has to be implemented with suitable procedures transforming the
theoretical quantities L (luminosity) and Te (effective temperature)
into reliable predictions about  magnitudes in selected photometric
bands.   To look into this problem Fig. \ref{atm} shows our 
Z=0.0002 12 Gyr isochrone (with diffusion) as transformed into the CM diagram 
according to  selected available choices about the color-temperature 
and magnitude luminosity transformations. One finds that 
the most recent evaluations given by Castelli et al. (1997 a,b)
are very similar to previous evaluations given by Buser 
\& Kurucz (1978, 1992) all along the TO region as well as in the 
redder portion of the HB, whereas Kurucz (1992) gives slightly redder
TO colors. In all cases one finds that the difference affects
mainly the colors, whereas bolometric corrections appear remarkably
similar. Thus V magnitudes appears independent of the choice of the models.   

If not otherwise advised, in this paper we will rely on the 
model atmospheres by Castelli et al. (1997 a,b). 
Here we note that preliminary computations (Castelli, private communication)
for the case Z=0.0002 have already shown that the model 
atmospheres can be further improved. As a matter of fact, 
one finds that when both the new solar abundances and the enhancement
of $\alpha$-elements is taken into account, model atmospheres 
for metal poor stars ([Fe/H]= -2.0) reach a better  
agreement with available empirical estimate by Alonso et al. (1996). 
Fortunately, the same Fig. \ref{atm} shows that such improvement affects only
the slope of the MS at the lower luminosity, leaving in particular
unaffected the predicted magnitude of candidate candle stars.

\begin{figure}[htbp]
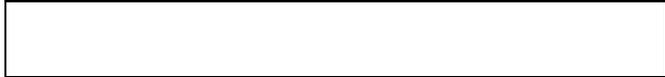
   
\picplace{1cm}
\caption{  The 9, 12  and 15 Gyr isochrones as computed under the labeled 
assumptions about the mixing length parameter and translated in the
CM diagram according to Castelli et al. (1997 a,b). }
\label{mix}
\end{figure}

As is well known,  MS models are also affected
by an  intrinsic uncertainty  due to the assumptions about the 
value of the mixing length parameter. The effect of varying the
adopted mixing length within the  (reasonable) interval l= 1.3-2.0 Hp
is shown in Fig. \ref{mix}, where we report selected isochrones for the 
labeled values of ages and mixing length parameters. One finds that
for ages ranging from 9 to 15 Gyr the MS color at Mv=6.0 decreases 
by about  $\Delta$(B-V)$\sim$0.02 mag, whereas
decreasing the mixing length from 2.0 down to 1.3 Hp (the local pressure
scale height) gives
again a decrease of this color of about $\Delta$(B-V)$\sim$0.03 mag, 
depending on the cluster age. Thus the color of the MS at Mv=6.0
is far from being a firm prediction of theory. The same
figure shows that the magnitude of the MS at B-V=0.55 (i.e. around
theoretical predictions for Mv=6) for a given cluster metallicity
can move by about 0.28 mag.($\sim$ 0.18 mag. for a variation 
from 1.3 to 2.0 Hp and $\sim$ 0.1 mag. for a difference in age from 9 to 15 Gyr),
preventing the use of theoretical MS models as precise distance calibrators.

When moving toward more advanced evolutionary phases, as already 
noticed by Chaboyer (1995), the calibration of TO 
magnitudes in terms of cluster ages does depend on the assumptions
about the mixing length parameter. One finds that such a dependence 
increases when the cluster ages and/or the metallicity is increased,
for the simple reason that in both cases  TO stars become cooler 
and, thus, more affected by external convection. As a result, the most metal
poor and younger TO (i.e., the hottest TO) are barely affected 
by the treatment of convection all over the explored range of
mixing length parameters (1.3 $\div$ 2.0 Hp). 
As an example, for Z=0.0002 and for the age t= 9 Gyr, a change from l=
1.3 to 2.0 Hp moves the TO magnitude by only 0.02 mag. However, for the 
same assumed metallicity, at t=14 Gyr, the same variation of the mixing length parameter
gives a difference in the TO magnitude of the order of 0.1 mag.\\ 

According to the  quoted evidences, we will proceed in the 
discussion of theoretical predictions with the warning that
discussed magnitudes should be reliable, whereas colors are
affected by the discussed uncertainties  in both stellar and 
atmospheric models. Detailed data for the computed isochrones 
are available by anonymous ftp at astr18pi.difi.unipi.it (/pub/globular).

In the next section we  discuss  low mass stars
during the major phase of H burning, namely from the initial main
sequence to the subgiant phase, just after the exhaustion of
central H. We selected this evolutionary phase according to the evidence 
that the corresponding stellar models rely on the "minimum" input
physics  as given by an equation of state (EOS) not too far from a
perfect gas, by radiative opacities and, finally,
by sufficiently well known H burning rates (see, e.g., Brocato et al. 1998)
Correspondingly, the quoted evolutionary phases
have to be regarded as the most solid predictions of theory. 
Section 4 will deal with a
discussion of more advanced evolutionary phases, from the red giant branch
to the AGB through the phase of central He burning, where more 
physics has to be added, like neutrino production by weak interactions
or the physical behavior of electron-degenerate
matter.
In both cases we  find that  theoretical predictions reach 
a good agreement with observational constraints. Following 
such an agreement, we  discuss 
relevant evolutionary features and, in particular, the estimate of cluster 
ages based on the HB luminosity level. A brief final discussion  closes the
paper.

\section{H burning phases.}

\begin{figure}[htbp]
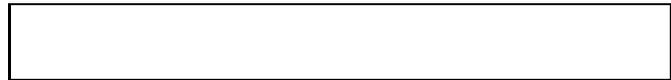
   
\picplace{1cm}
\caption{Comparison of present isochrones with Hipparcos subdwarf absolute
magnitudes (Gratton et al. 1997). Each panel shows the run of the 10 and 12 
Gyr isochrones (full lines) for the given value of [Fe/H]. Triangles (and
the dashed line)
show the shift of the 10 Gyr isochrones caused by an enhancement 
[$\alpha$/Fe] $\simeq$ 0.3. As a comparison, dotted lines show the
run of 10 Gyr isochrones with different [Fe/H], as labeled. 
Filled squares and open squares
indicate single stars and detected or suspected binary stars,
respectively.}
\label{Hip}
\end{figure}

As a first test of theoretical results,  Fig. \ref{Hip} (a to c)
shows a comparison between  isochrones (with sedimentation) and Hipparcos data 
(Gratton et al. 1997, Chaboyer et al. 1998) for field subdwarfs in selected intervals of 
[Fe/H] values. Each panel compares  isochrones at t=10 and 12 Gyr and for
a given [Fe/H] value with  subdwarfs with  [Fe/H] estimates within
$\pm$ 0.1 dex. Triangles in each panel  show the
estimated shift in the isochrone if the suggested enhancement in
$\alpha$-elements  [$\alpha$/Fe]$\simeq$0.3 (Carretta \& Gratton 1997,
Gratton et al. 1998a) is accounted as an increase in the total value of Z. 
For the sake of comparison,  each panel shows also  the location 
of the MS loci but for different  metallicities. 

Inspection of the three panels shows  that $\alpha$-elements 
play a  minor role, since the corresponding shift of the isochrone 
lies  within the observational errors. Bearing in mind that
sedimentation does not affect the location of MS stars, one finds that
the agreement between theory and observation appears as fine as it
can be expected according to the standard errors in absolute 
magnitudes and the spread in metallicity. One finds that the popular Reid's
statement  "current models overestimate the change in luminosity 
with decreasing abundance for extreme metal poor subdwarf" is hardly
supported. In fact, the three  panels of Fig.\ref{Hip}
suggest that no contradiction
appears between extant theory and Hipparcos observations.

\begin{figure}[htbp]
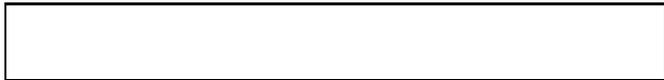
   
\picplace{1cm}
\caption{Present predictions of B-V colors at Mv=6.0
for the various metallicities compared with observational data 
given by Gratton et al. (1997a) on the basis of Hipparcos observations.}
\label{MvHip}
\end{figure}

Figure \ref{MvHip} shows the comparison between  present
predictions of B-V colors at Mv= 6.0 and Hipparcos observational data as
presented by Gratton et al. (1997). One finds that theoretical 
data appear in satisfactory agreement with observations, whose error
appears of the same order of magnitude  of the theoretical uncertainty
discussed in the previous section. The best fit 
to the theoretical data gives:\\

\noindent
(B-V)= 0.890 + 0.256 [Fe/H] + 0.048 [Fe/H]$^2$\\
\noindent
(at Mv=6.0)\\

\noindent
which practically overlaps the  relation given by Gratton et al. (1997a)
 except for a difference in the zero point by 
$\Delta$(B-V)= 0.014. As an alternative approach to the effect of
metallicity one finds for the MS magnitude at (B-V)=0.6:\\

\noindent
Mv= 3.889 - 2.160[Fe/H] - 0.509[Fe/H]$^2$\\
\noindent   
(at (B-V)= 0.6)\\

\noindent
One can use either of these two relations for correcting the metallicity 
effects in data in Fig.3, by shifting 
the observational data to the [Fe/H] value of
the computed isochrone. However, one finds that such a procedure
does not improve the fitting given in Fig. \ref{Hip}, as an evidence
that parallax errors  (i.e., errors in absolute magnitudes) overcome
the effect of metallicity in the chosen dwarf samples.

\begin{figure}[htbp]
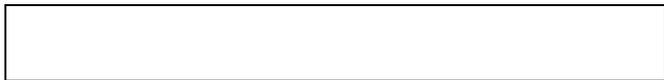
   
\picplace{1cm}
\caption{Visual magnitude of the TO isochrones as a function of the
logarithm of the age, for the labeled metallicities for models with 
(solid line) and without (dashed line) element diffusion.}
\label{TOt}
\end{figure}

\begin{figure}[htbp]
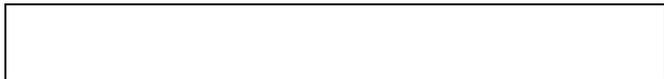
   
\picplace{1cm}
\caption{TO visual magnitudes  for t=10 Gyr as a
function of the metallicity for present models as compared 
with similar results available in the current literature.
For DCM 1997 models CM indicates the adoption by the authors
of the Canuto \& Mazzitelli (1991) treatment of overadiabatic 
convection while MLT indicates the adoption of the usual mixing length
theory.}
\label{TOZ}
\end{figure}

When moving from the MS to more advanced phases, Fig. \ref{TOt} gives 
TO V-magnitudes as a function of age for
the various adopted metallicities. One finds that
for each given metallicity the TO magnitudes can be arranged as a
linear function of the logarithm of the age, with a maximum error (for Z=0.001) not exceeding
0.4 Gyr. Table 1 gives the coefficients of the linear relations
connecting the TO visual magnitudes with  cluster ages (in Gyr) for isochrones with element
diffusion taken into account.  Figure \ref{TOZ} shows the comparison
between current  predictions for
the TO magnitudes at t=10 Gyr and for the four adopted metallicities
with similar results in the literature. 
As already predicted in paper I, one finds that present evaluations
tend to decrease the cluster age for any given TO magnitude, the
differences in Fig. \ref{TOZ} being mainly the consequence
of the difference already discussed in Paper I concerning theoretical
luminosities. In passing we note  that at the larger metallicities present 
magnitudes do not
overlap Mazzitelli et al. (1995) predictions, as occurred for
luminosities. This is mainly due to a small difference in adopted visual
magnitude of the Sun: MDC adopted Mv$_\odot$=4.79 mag. against our preferred 
canonical value Mv$_\odot$=4.82 mag.  

\begin{figure}[htbp]
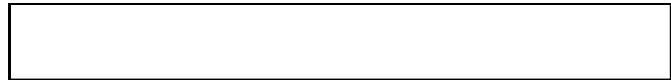
   
\picplace{1cm}
\caption{Visual magnitude at the isochrone BTO for present models
with diffusion as a function of the logarithm of the age, for the labeled metallicities.}
\label{BTO}
\end{figure}

As already known, the TO magnitude, which is defined as 
the magnitude of the bluest point of the isochrone, is 
a difficult observational parameter in particular in the most
metal poor clusters. Moreover, small variations in the shape of the 
theoretical isochrone can produce relevant variation in the magnitude
of the nominal TO. This explains the small departures from 
linearity  disclosed by data plotted  in Fig. \ref{TOt}.
In a recent paper Chaboyer et al. (1996) suggested a different calibrator of
the globular cluster age, as given by the visual magnitude of the so called
BTO point, that is the point  brighter than the TO and redder by
$\Delta$(B-V)=0.05.
The main advantage of  Mv(BTO)  is that this 
point is not in the vertical turn-off region and thus it is easier 
found in the observed color magnitude diagram.
Figure \ref{BTO} shows that the visual magnitude of theoretical BTO 
appears much more regularly depending on the cluster age than  Mv(TO)
does, further supporting  the use of such a parameter.

\section{Horizontal branches.}

As a preliminary point, let us recall that in Paper I HB models have
been computed assuming initial He core masses from a 0.8 M$_{\odot}$
progenitor. Strictly speaking, according to data given in the same 
paper, this implies an age of the
order of 11 Gyr for the most metal poor cluster, increasing up to
16 Gyr for Z=0.006.  In Paper I we already discussed this issue,
reaching the conclusion  that the expected variation of HB luminosity can be neglected
over a reasonable range of ages. Now we add that detailed numerical
experiments show that a variation of the cluster ages of $\pm$ 5 Gyr
gives a variation in the predicted HB magnitude of $\pm$ 0.02 mag. 
According to the calibration given 
in Fig. \ref{TOt} a similar error in the cluster distance modulus
would produce an error in the cluster age not larger than 0.3 Gyr which,
however, could be easily taken into account whenever a better 
precision would be required. 
  
\begin{figure}[htbp]
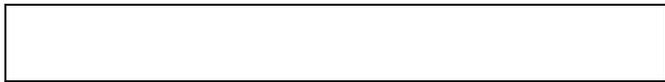
  
\picplace{1cm}
\caption{Visual magnitude of the ZAHB at logTe=3.85 as a function of
the metallicity for present models with (solid line)
and without diffusion (dashed line) compared with values for the same quantity available in the
literature. CCP91 indicates models by Castellani, Chieffi \& Pulone (1991).}
\label{ZAHBZ}
\end{figure}

Table 2 gives details 
for HB structures in the observational plane at the various metallicities.
Selected evolutionary quantities for the new HB models with Z=0.0006 
are given in the Appendix of this paper.
A comparison of theoretical predictions for HB with Hipparcos data
has been already presented in Paper I. Figure \ref{ZAHBZ}
compares the magnitude of the ZAHB (at logTe=3.85) with previous
evaluations on the matter. One recognizes the
not negligible increase in the predicted HB luminosity induced
by the improved physics as well as the fair agreement, toward the lower 
metallicities, with the recent computations by Caloi et al.
(1997). As already predicted (Castellani et al. 1991) one finds that
the dependence of the ZAHB luminosity on the metallicity increases
with metallicity. The best fit of data in Fig. \ref{ZAHBZ} gives:\\

\noindent
Mv = 0.993 + 0.461[Fe/H] + 0.087[Fe/H]$^2$\\

\noindent
which reproduces the theoretical predictions for ZAHB with diffusion 
within less than 0.01 mag all over the assumed range of metallicities. 
If diffusion is not taken
into account the above magnitude has to be decreased by about  0.04 mag.

For metallicities lower than Z=0.001 the above relation can be approximated
by a linear relation, as usually adopted in the literature. We find:

\noindent
Mv=  0.18[Fe/H] + 0.74 \hspace{2cm} (no diffusion)\\
Mv=  0.18[Fe/H] + 0.77 \hspace{2cm} (diffusion).\\

\noindent
These results can be usefully compared with the fairly large amount
of observational relations presented in the literature:\\

\noindent
Mv=  0.15[Fe/H] + 1.01   \hspace{1.3cm} (Carney et al. 1992)\\
Mv=  0.15[Fe/H] + 0.73   \hspace{1.3cm} (Walker 1992)\\
Mv=  0.15[Fe/H] + 0.84   \hspace{1.3cm} (De Santis 1996)\\
Mv=  0.18[Fe/H] + 0.74   \hspace{1.3cm} (Gratton et al. 1998b)\\
Mv=  0.19[Fe/H] + 0.97   \hspace{1.3cm} (Clementini et al. 1995)\\
Mv=  0.23[Fe/H] + 0.83   \hspace{1.3cm} (Chaboyer et al. 1998)\\
Mv=  0.30[Fe/H] + 0.94   \hspace{1.3cm} (Sandage 1993)\\
  
\noindent
It appears that present predictions show a dependence on the metallicity
in reasonable 
agreement with the evaluation by Walker (1992) and the more recent 
evaluation given by Gratton et al. (1998b) based on Hipparcos results, and
definitely smaller than required by Sandage in his scenario for
explaining the Oosterhoff dichotomy. 
However, before comparing the zero point of the magnitudes
one has to recall
that our previous relations refer to the ZAHB luminosity level, whereas
observational data refer to the mean luminosity of the HB at the color
of the RR Lyrae Gap. The connection between the two luminosity level has been
already discussed in several paper (see Caputo et al. 1987, Carney et al. 1992, 
Cassisi \& Salaris 1997). One can
safely assume $\Delta$Mv=0.08 mag as a suitable estimate of the 
difference
in magnitude between RR and ZAHB. Thus our zero points will become 
Mv=0.69 (diffusion) or Mv=0.66 (no diffusion), respectively. One finds that 
we are predicting HB  with the same dependence on metal content and
only 0.05 mag. brighter with respect to the recent estimates
by Gratton et al. (1997).

\begin{figure}[htbp]
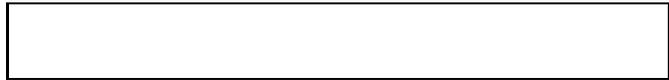
   
\picplace{1cm}
\caption{The difference in magnitude between HB and TO as a function
of the age for present models with (solid line) and without (dashed line)
element diffusion for the labeled values of metallicities.}
\label{TOHB}
\end{figure}

One can finally connect theoretical results concerning HB stars
with previous predictions about the TO magnitudes to give a theoretical
calibration of the difference in magnitude between HB and TO, $\Delta$V(TO-HB),
 often used as age indicator for galactic globulars. This is shown in Fig. \ref{TOHB},
where we report $\Delta$V(TO-HB)  as a function of the
age for the four selected assumptions about stellar metallicity.  

\section{The role of the original He content.}

All the theoretical
computations referred in the previous sections assume an amount of
original He content equal to Y=0.23, which is, at the present, the most
popular and widely adopted estimate for the original He abundance in metal
poor, Population II stars. However, in Paper I we already discussed
the large uncertainty in the evolutionary determination of this
parameter. According to such an evidence, we will investigate the
role played by the assumption about Y on the present theoretical
scenario.  To this purpose we reinvestigated the evolutionary behavior of
the Z=0.001 models (with element diffusion) but under the two 
alternative assumptions  Y=0.21 or Y=0.25. 

 \begin{figure}[htbp]
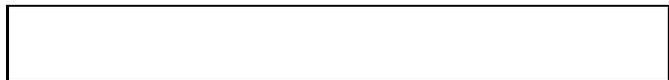
   
\picplace{1cm}
\caption{Present isochrones in the CM diagram for Z=0.001,
and the labeled values of age and original He abundances.} 
\label{MSY}
\end{figure}

Figure \ref{MSY} compares selected isochrones for Z=0.001 and
for the three adopted values of helium abundance.
As a result one finds the following   relations:\\

\noindent
unevolved  MS (at B-V=0.6):\\  
\hspace{1.0cm} $\Delta$Mv/$\Delta$Y$\sim$2.5  (slowly depending on (B-V)\\

\noindent
TO (at t=10 Gyr):\\  
\hspace{1.0cm} $\Delta$Mv= 0.019 (Y=0.21$\div$0.23)\\  
\hspace{1.0cm} $\Delta$Mv=0.034  (Y=0.23$\div$0.25)\\

\noindent
HB (at logTe=3.85):\\
\hspace{1.0cm} $\Delta$Mv= -0.084 (Y=0.21$\div$0.23)\\
\hspace{1.0cm} $\Delta$Mv= -0.068 (Y=0.23$\div$0.25)\\

As a whole, present results (where element sedimentation is taken into
account) appear in reasonable concordance with
canonical evaluations given by Renzini (1991) for the TO magnitudes and
by  Buzzoni et al. (1983) or, more recently, by  Bono et al. 
(1995) for HB luminosities. 
One finds that a variation of Y within the assumed
limits (Y=0.23$\pm$0.02) gives a maximum variation of 0.034 mag. in the TO magnitude and
of 0.10 mag. in the difference of magnitude between TO and HB. According to the
calibration given in previous sections this uncertainty implies 
an error of about 
0.2 Gyr for ages from the TO magnitude, and an error of about 1 Gyr
for ages from the "vertical" method, i.e., from the difference in magnitude 
between TO and HB.

From the previous sections one also derives $\Delta$Mv(TO)$\sim$ 0.37 $\Delta$[Fe/H]
and $\Delta$Mv=0.18 $\Delta$[Fe/H]. Assuming an uncertainty of $\pm$ 0.2 dex on
current evaluations of cluster metallicity, we obtain  that such
an error drives, by itself, a variation of, about, 0.08 mag in
the predicted TO, and of about 0.04 mag on the difference in
magnitude between TO and HB. As a result, one finds that with
the assumed uncertainty in both Helium and metal content, \underline{even with
perfect photometry}, the age of
a cluster cannot be determined better than $\pm$ about 0.8 Gyr
from the TO magnitude or better than, about, 1.4 Gyr from the
vertical method.

 We remind that in all cases we assumed scaled solar composition
as given by Grevesse \& Noels (1993). However, for metallicity not too much high,
the effect of the $\alpha$-enhancement on the evolutionary tracks
and isochrones can be simulated by using a scaled solar mixture of total metallicity
equal to the actual one (see Salaris et al. 1993). Detailed calculations 
recently performed by Salaris \& Weiss (1998b) show that the $\alpha$-enhancement
does not influence at all the $\Delta$V(TO-HB) parameter for metallicities up to
Z$\approx$0.004, while for Z=0.01 the variation of  $\Delta$V(TO-HB) is about 0.1 mag.
Thus the influence of the $\alpha$-enhancement on  $\Delta$V(TO-HB) is expected to be
negligible up to our highest adopted metallicity (Z=0.006). In the case of 
$\alpha$-enhancement we only expect, for Z=0.006, a shift of the color
of the RG branch, well within the known uncertainties due to the uncertainty
on the efficiency of superadiabatic convection and on the adopted models
atmospheres.

\begin{figure}[htbp]
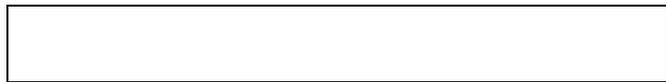
  
\picplace{1cm}
\caption{Isochrones for ages between 10 and 12 Gyr and ZAHB compared
to the CMD of M68 (data from Walker 1994). Composition, distance modulus and reddening 
used for the fit as labeled. The adopted mixing length is 2.0 Hp. }

\label{M68}
\end{figure}

\begin{figure}[htbp]
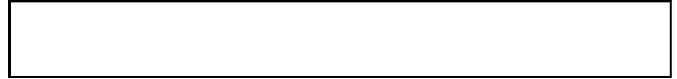
   
\picplace{1cm}
\caption{Isochrones for age between 9 and 11 Gyr and ZAHB compared to the
CMD of M5 (data from Sandquist et al. 1996). Composition, distance modulus and reddening
as labeled. The adopted mixing length is 2.3 Hp.}
\label{M5}
\end{figure}

As already discussed  by Caputo et al. (1983)
one finally finds that MS and HB have
opposite behavior as far as the amount of original He is
concerned. This implies that the difference in magnitude between the
HB and the MS at a given color is in principle an indicator of the
original He content independent on the cluster age and not affected,
as the well known parameter R is, by the still large inaccuracy on
the rate of the $^{12}$C-$\alpha$ reactions. This indicator  is becoming
more and more relevant vis-a-vis the increased capability of precise
photometry of faint MS stars in galactic globulars. Combining the previous
relations one finds $\Delta$Mv(HB-MS)$\sim$6.5$\Delta$Y. 
Thus $\Delta$Mv(HB-MS) appears rather sensitive to the amount of
original He, increasing by 0.07 mag. when Y is increased by only 0.01.
However one should remember that  
significant theoretical uncertainties, due to the
uncertainty in the efficiency of superadiabatic
convection and to the choice of color transformations, 
together with observational uncertainties, due 
to the evaluation of the cluster reddening and metallicity,
make de facto this quantity a difficult parameter to be 
used as helium calibrator.

\begin{figure}[htbp]
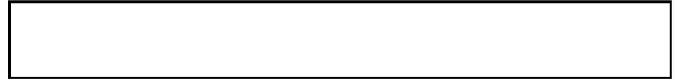
   
\picplace{1cm}
\caption{ Theoretical predictions concerning the HB magnitudes
from the present paper (full line: diffusion) or from CCP (dashed line)
compared with recent observational estimates as derived either from
the subdwarf fitting (open symbols) or from RR Lyrae parallaxes
(filled symbols).}

\label{stat}
\end{figure}

\section {Discussion and final remarks}

Before closing the paper let us use the presented theoretical
scenario to fit the C-M diagram of selected globular clusters. For a
first test we chose C-M diagram presented by Walker (1994) for the
metal poor cluster M68. A preliminary discussion on that matter has
been recently presented by Brocato et al. (1997) on the basis of Paper
I evolutionary results and using Kurucz (1992) transformations. Without
repeating the discussion given in that paper we present in Fig. \ref{M68}
 the best fitting of the observational diagram as obtained on
the basis of our Z=0.0002 isochrones with element diffusion, adopting the 
very last
version of Kurucz's model atmosphere (Castelli 1997a,b), where the new solar abundances
and the enhancement of $\alpha$-elements have been taken into account.

The best fitting is achieved for an age of about 11 Gyr, with distance modulus
and reddening as labeled. Note that to fit the color of the RG branch we
adopted l= 2.0Hp. However, Fig. \ref{M68} shows that with such an
assumption the MS appears slight  bluer than observed, requiring lower  
values of the mixing length parameter and indicating that the CM
diagram of the cluster can hardly be fitted with precision assuming a constant
mixing length for all the evolutionary phases. By neglecting this
(false) problem of temperature of cool stars, one finds that the
best fitting requires a cluster distance modulus DM= 15.30, in 
excellent agreement with Gratton et al. (1997) who used Hipparcos 
parallaxes to find for the cluster DM=15.31 and a (mean) age of 11.3 Gyr. 

As a test of theory at larger metallicity, Fig. \ref{M5} present the
best fitting of the C-M diagram presented by Sandquist et al. (1996)
for the intermediate metallicity cluster M5. Now the best fitting gives
an ages of about 10 Gyr and a distance modulus DM= 14.54. On the basis of
Hipparcos parallaxes Gratton et al. (1997) give DM=14.60$\pm$0.07 with a 
(mean) age of 10.5 Gyr.  On the same ground, Chaboyer et al. (1998) gives 
DM= 14.51$\pm$0.09 with an age of 8.9$\pm$1.1 Gyr.

The presented evolutionary models appears
in both cases to give excellent agreement with independent evaluations of
the cluster distance moduli based on Hipparcos data. This strongly increases
the confidence in HB stars as (theoretical) standard candles and, in the 
same time, in the reliability of the derived cluster ages. One may notice that 
both clusters we are dealing with have been recently fitted with
less updated physics, corresponding to step 4 in Paper I. In that case
it was derived t= 12.2 Gyr for M68 (Salaris et al. 1997) and, with
slight different assumptions about the cluster chemical composition (Y=0.235, 
Z=0.0015), t= 10.9 Gyr for M5 (Salaris \& Weiss 1998a). Comparison 
with present results casts new light on the further rejuvenation of cluster
ages induced by both the subsequent updating of the physics and 
the introduction of the element diffusion.

One can safely assume a conservative error of $\pm$ 0.1 mag 
in our estimates of the difference in magnitude between TO and HB,
due to the arbitrarity introduced when the theoretical HB and isochrones
are fitted to an observational CM diagram. As an example, we note that Brocato et al. 1997
using the theoretical models of paper I, find for M68 a distance modulus (DM=15.25) slightly
different from the present result, due to the slightly different way
in which the theoretical HB is fitted to the observational one.
This variation in the distance modulus together with the adopted
color transformations (which influence the look of the fit) yields
a difference of $\sim$ 1 Gyr in the estimated age.  
With the above quoted assumptions about the uncertainty in the chemical
composition we conclude that the adopted theoretical scenario gives
for our clusters:

\noindent
M68: t= 11 $\pm$1.4 $\pm$1.0 Gyr\\
\noindent
M5: ~t= 10 $\pm$1.4 $\pm$1.0 Gyr\\

\noindent
where the first error is due to the uncertainty in the chemical composition,
while the second represents the uncertainty in the fit.
In passing, we note that the close similarity
between the C-M diagrams of M68 and M92 (as, e.g., discussed in 
Brocato et al. 1997 and Salaris et al. 1997) drives to the conclusion that both
clusters should have quite similar ages. According to the above
discussion we suggest for these very metal poor globulars an age
of the order of 11 Gyr against the 14 Gyr
recently derived by Pont et al. (1998).  One may notice that the above
age evaluations could suggest a possible correlation between
cluster ages and metallicity, the more metallic cluster being 
also the younger one. The evidence from the figure appears clear enough,
however the source of possible errors in the photometry and in the 
fitting do not allow firm conclusions about
a problem which deserves much more accurate investigations.

As a final remark, let us here remind once more that the above age
estimates rely on the theoretically predicted HB luminosity. We have
already quoted the good agreement of such a prediction with cluster
distance moduli as derived by the fitting of Hipparcos subdwarf
magnitudes. However one has also to remind that several estimates of RR Lyrae
luminosities based on Hipparcos data give sensitively fainter 
magnitudes. As shown in Fig.\ref{stat}, the issue is far from being 
clearly settled. Here we can only say that if such faint magnitudes
will be eventually confirmed,  present theory is obviously overestimating
the He cores at the end of RG evolution, likely as a consequence
of a corresponding overestimate of the efficiency of cooling along
the RG phase. In this case, data in Table 1 indicate that cluster ages
should be increased by $\Delta$logt$\sim$0.4 $\delta$Mv, where   
$\delta$Mv represents the difference between the actual and the 
predicted magnitudes of RR Lyrae stars.

\section{Acknowledgments}
It is a pleasure to thank Giuseppe Bono for a critical reading of
the manuscript and for valuable suggestions. One of the authors, S.C.,
acknowledges the grant from C.N.A.A.

\section{Appendix}

Table 3 lists selected evolutionary quantities for 
the new models with Z=0.0006
when the efficiency of element sedimentation is neglected.
Left to right one finds:  mass of the model, age, luminosity and effective
temperature at the track Turn-Off (TO),  mean luminosity 
of the RGB ``bump'',
age at the He flash and the corresponding luminosity, mass of the He 
core and amount of extra-He brought to the surface at this
time. Table 4 gives the same quantities but for models 
where element sedimentation is taken into account according to the procedure
reported in Paper I. 
Table 5 gives details on the isochrone TO over the explored range of ages
for the new models with Z=0.0006.
Table 6 gives selected evolutionary quantities for the new HB models with Z=0.0006.\\
Finally we compare the evolutionary characteristics of our models
with those of similar models, for the same chemical composition and slightly
different physics, recently presented by Straniero et al. (1997)
finding a substantial agreement.  Regarding the H burning phases,
there is a good agreement in the luminosity and effective temperature
of the TO region, while our TO ages are lower by about 10\%, presumably
due to the adoption of a different EOS. As far as the evolutionary characteristics 
at the He flash is concerned,
the value of the extrahelium brought to the surface during the first dredge-up is
the same, while our He cores are larger of about 0.005 M$_{\odot}$. This very slight difference
could be due to our adoption of updated calculations for plasma
neutrino energy losses (Haft et al. 1994).


\begin{thebibliography}{9}

\bibitem{Alon} Alonso A., Arribas S., Martinez-Roger C., 1996, A\&A 313, 873
\bibitem{} Bono G., Castellani V., Degl'Innocenti S., Pulone L. 1995, A\&A 297,115
\bibitem{Bono} Bono G., Caputo F.,  Castellani V.,  Marconi M., 1997, A\&AS 121, 327 
\bibitem{Bro} Brocato E., Castellani V., Piersimoni A., 1997, ApJ, 491, 789
\bibitem{Bro2} Brocato E., Castellani V., Villante F., 1998, MNRAS, in press
\bibitem{BK78} Buser R., Kurucz R. L., 1978, A\&A 70, 555 
\bibitem{BK82} Buser R., Kurucz R. L., 1992, A\&A 264, 557 
\bibitem{Buz} Buzzoni A., Fusi Pecci F.,Buonanno R., Corsi C.E. 1983, A\&A 128, 94
\bibitem{Caloi} Caloi V., D'Antona F., Mazzitelli I. 1997, A\&A 320, 823
\bibitem{Can} Canuto F., Mazzitelli I., 1991, ApJ 370, 295
\bibitem{Cap} Caputo F., Cayrel R., Cayrel de Strobel G. 1983, A\&A 123, 135
\bibitem{Capu} Caputo F., Martinez Roger C., Paez E., 1987, A\&A 183, 228
\bibitem{Carney92} Carney B.W., Storm J.J., Rodney V., 1992, ApJ 386, 663
\bibitem{Car}Carretta E., Gratton R.G., 1997 A\&AS 121,95
\bibitem{CASSal} Cassisi S., Salaris M., 1997, MNRAS 285, 593 
\bibitem{Cass} Cassisi S., Castellani V., Degl'Innocenti S., Weiss A., 1998, A\&AS 129, 267
\bibitem{Cqu} Castellani V., Quarta M.L.1987,  A\&AS, 71,1
\bibitem{CCP91} Castellani V., Chieffi S., Pulone L., 1991, ApJS 76, 911
\bibitem{Cast} Castelli F., Gratton R.G., Kurucz R.L., 1997a, A\&A 318, 841
\bibitem{Cast} Castelli F., Gratton R.G., Kurucz R.L., 1997b, A\&A 324, 432
\bibitem{CHAB95} Chaboyer B. 1995, ApJ 444, L9
\bibitem{Cha} Chaboyer B., Demarque P., Kernan P.J., Krauss L.M., Sarajedini A., 1996, MNRAS 283, 683 
\bibitem{Cha97} Chaboyer B., Demarque P., Kernan P., Krauss L.M., 1998, ApJ 494, 96
\bibitem{Cle} Clementini G., Carretta E., Gratton R.G. et al., 1995, AJ 110, 2319
\bibitem{Dantona} D'Antona F., Caloi V., Mazzitelli I. 1997, ApJ 477, 519
\bibitem{DeSan} De Santis R. 1996, A\&A 306, 755
\bibitem{Feast} Feast M.W., Catchpole R.M., 1997, MNRAS 286, L1 
\bibitem{Fern} Fernley J., Barnes T.G., Skillen I. et al., 1998, A\&A 330, 515 (1998) 
\bibitem{Grat0} Gratton R. G., 1998 preprint astro-ph/9710271 to appear on MNRAS
\bibitem{Grat1} Gratton R.G., Fusi Pecci F. Carretta E., Clementini G., Corsi C.E., Lattanzi M.G., 1997, 
ApJ 491, 749
\bibitem{Grat2} Gratton R.G., Carretta E., Castelli F., 1998a, A\&A in press
\bibitem{Grat3} Gratton R.G., Clementini G., Fusi Pecci F., Carretta E., 1998b, in `` Views on
distance indicators'', F. Caputo ed. Mem. SAIt in press
\bibitem{Gre} Grevesse N., Noels A., 1993 in ``Origin and evolution of the elements'', ed. N. Prantzos, E. Vangioni-Flam,
 M. Cass\'e, (Cambridge University Press, Cambridge), p.15
\bibitem {Haft} Haft M., Raffelt G., Weiss A., 1994, ApJ 425, 222
\bibitem{Kur92} Kurucz R.L. 1992, in IAU Symp. 149, The stellar Population of Galaxies,
eds. B. Barbuy, A. Renzini (Dordrecht: Kluwer), 225
\bibitem{Layden} Layden A.C., Hanson R.B., Hawley S. L., Klemola A.R., Hanley C.J. 1996, AJ 112, 2110
\bibitem{MDC95} Mazzitelli I., D'Antona F., Caloi V., 1995, A\&A 302, 382 
\bibitem{Pont} Pont F., Mayor M., Turan C., VandenBerg D.A.,  1998, A\&A  329, 87 
\bibitem{Popow} Popowski P., Gould A. preprint astro-ph/9802168
\bibitem{Reid} Reid I. N., 1997, AJ 114, 161 
\bibitem{} Renzini A. 1991, in ``Observational tests of cosmological inflation'', ed. T.Shanks et al. (Dordrecht: Kluwer), p.7
\bibitem{SalW} Salaris M., Weiss A., 1998a, A\&A, in press
\bibitem{SalW} Salaris M., Weiss A., 1998b, A\&A, submitted
\bibitem{Sal93} Salaris M., Chieffi S., Straniero O., 1993, ApJ 414, 580
\bibitem{Sal97} Salaris M., Degl'Innocenti S., Weiss A., 1997, ApJ 479, 665
\bibitem{Sand} Sandage A., 1993, ApJ 106, 703
\bibitem{Sandq} Sandquist E.L., Bolte M., Stetson,P.B.,Hesser J.E., 1996, ApJ 470, 910
\bibitem{} Straniero O., Chieffi A., Limongi M., 1997 ApJ 490, 425
\bibitem{Tsuj} Tsujimoto T., Miyamoto M., Yoshii Y., 1998, ApJ 492, L79
\bibitem{} VandenBerg D.A., Bolte M., Stetson P.B., 1996, ARA\&A 34, 461
\bibitem{Walk92} Walker A.R., 1992, ApJ 390, L81
\bibitem{Walk94} Walker A.R., 1994, AJ 108, 555

\end{thebibliography}
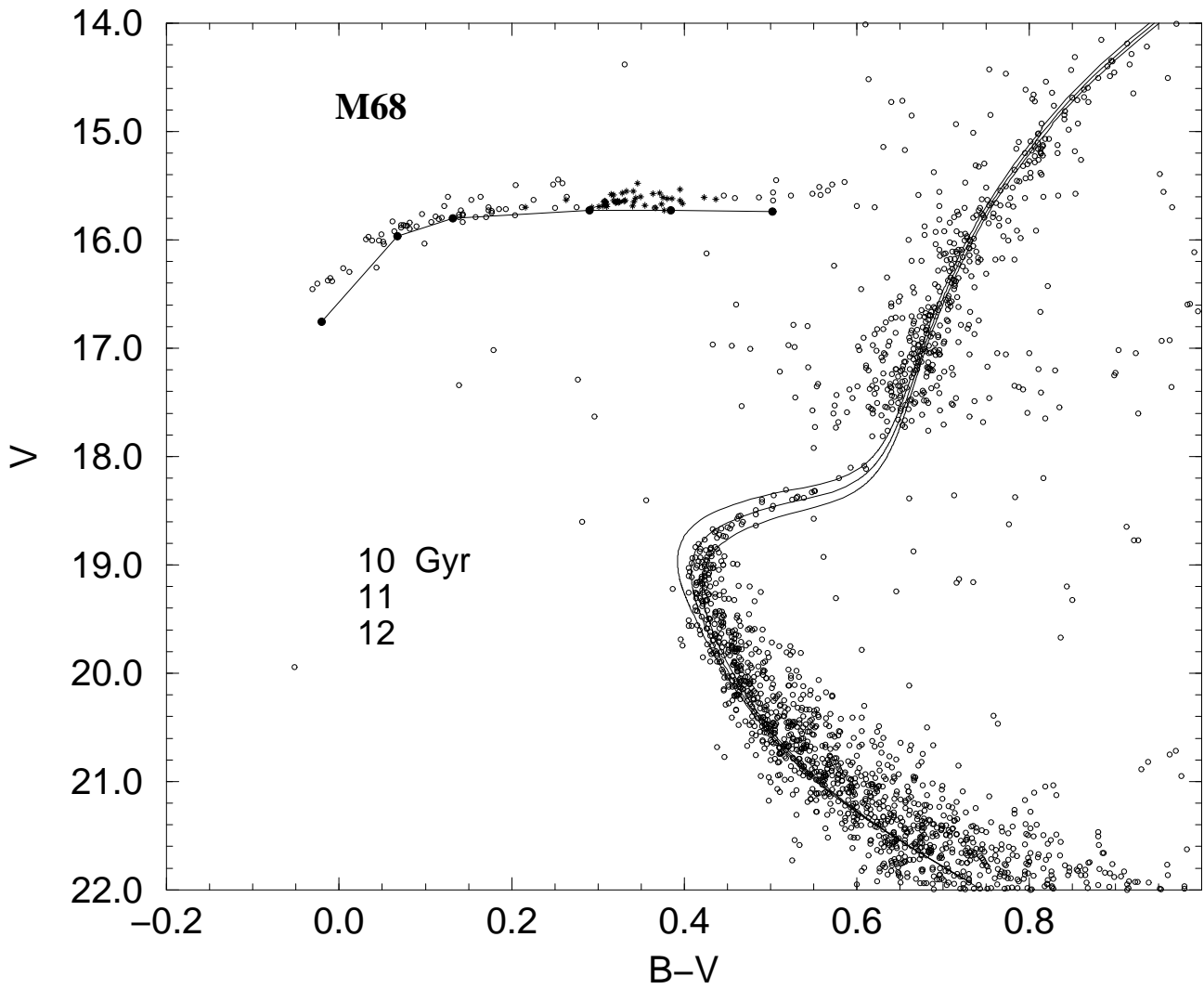
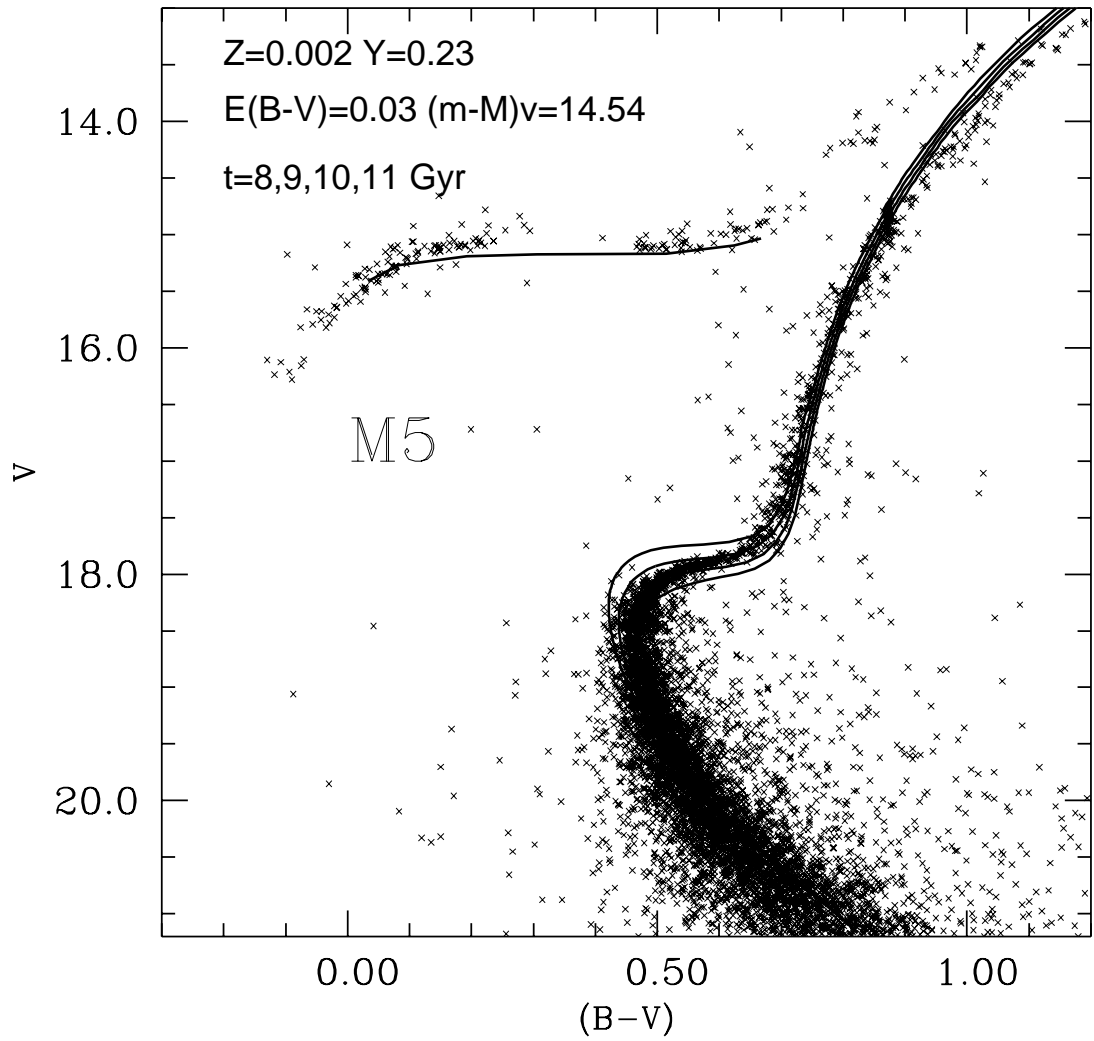
\end{document}